\documentclass[12pt, prd, showpacs]{revtex4}
\usepackage{amssymb}
\usepackage{amsmath}

\input{tcilatex}

\begin{document}

\title{Super-Penrose process for nonextremal black holes}
\author{O. B. Zaslavskii}
\affiliation{Institute of Astronomy, Kharkov V.N. Karazin National University, 4 Svoboda
Square, Kharkov 61022, Ukraine}
\affiliation{Institute of Mathematics and Mechanics, Kazan Federal University, 18
Kremlyovskaya St., Kazan 420008, Russia}
\email{zaslav@ukr.net }

\begin{abstract}
We consider particle collisions in the background of nonextremal spherically
symmetric static black holes. It is shown that debris of collision can have
indefinitely large energy at infinity, i.e. the super-Penrose process (SPP)
can occur. This property is sharply contrasted with that of rotating black
holes for which it is already established that the SPP is forbidden. The
Reissner-Nordstro\"{m} black hole serves as an example. If an external
central force exerts on particles, even the Schwarzschild background is
suitable for the SPP.
\end{abstract}

\maketitle

\section{Introduction}

The interest to high energy processes near black holes increased
significantly after the work \cite{ban}. It was shown there that if two
particles move towards the Kerr extremal black hole and collide in its
vicinity, the energy $E_{c.m.}$ in their center of mass frame can become
unbounded, provided one of two particle (called critical) has fine-tuned
parameters. This is called the Ba\~{n}ados-Silk-West (BSW) effect. The close
analogy of this effect exists also for extremal charged static black holes 
\cite{jl}. However, as far as the Killing energy $E$ of debris detected at
infinity is concerned, the situation differs radically for two
aforementioned cases. For rotating black holes, the energy $E$ of an
escaping particle at infinity is bounded \cite{p}, \cite{j}, \cite{z}.
Meanwhile, there is no such a bound for the extremal Reissner-Nordstro\"{m}
\ (RN) black hole. This was obtained in \cite{rn} and later confirmed in 
\cite{nem}. The process with unbounded $E$ at infinity is called the
super-Penrose process (SPP).

As far as nonextremal black holes is concerned, two problems existed here.
First, it was wide-spread belief that extremality is a necessary condition
for the BSW effect, so deviation from extremality weakens the effect \cite%
{berti}, \cite{ted}. However, it was shown in \cite{gp} that if instead of
one particle being exactly critical, a near-critical particle is used, and
deviation from the critical state is adjusted to the proximity of the point
of collision to the horizon in a special way, the effect survives. Moreover,
one can add a force acting on particles and this is consistent with the BSW
effect \cite{ne}. Second, it was unclear how to realize the BSW effect
physically. The most relevant situation corresponds to particles falling
from infinity. However, for rotating black holes, the centrifugal barrier
prevents the critical particle from reaching the nonextremal horizon \cite%
{gp} (see also case 2i in \cite{prd}, Sec.2 of \cite{gao} and \cite{kras}).
This can be repaired, provided additional constraints are imposed on the
scenario, because of which the turning point is situated closely to the
horizon \cite{can}.

However, there is an interesting question that, to the best of our
knowledge, was not posed up to now: whether or not the SPP is possible for
nonextremal black holes. It is considered in the present work. We show that
this is indeed possible. In this sense, there is a sharp contrast between
extremal and nonextremal black holes. One can think that this observation
may be useful for astrophysically relevant black holes since they are
nonextremal. It possesses some universal features in what any particles
moving in the background of a nonextremal black hole (even in the
Schwarzschild metric) and experiencing the action of some force can exhibit
this effect.

It is worth stressing that one should not confuse two different effects
connected with two different kinds of energy. The possibility of unbounded $%
E_{c.m.}$ (the BSW effect) when a force acts on particles moving near a
nonextremal black hole, was shown in \cite{F} that extended the results of 
\cite{jl}. In principle, the existence of the BSW effect is not sufficient,
in general, for the SPP, as is mentioned above. Therefore, the fact that the
BSW effect is possible in scenarios considered in \cite{can}, by itself does
not guarantee for them the SPP. However, we will show that this is indeed
the case for the situation under discussion that includes, in particular,
scenarios of \cite{can}. To the best of my knowledge, this is a first
example of considering the SPP for nonextremal black holes. In what follows,
we use the geometric system of units in which fundamental constants $G=c=1$.

\section{Basic formulas}

Let us consider the black hole metric%
\begin{equation}
ds^{2}=-N^{2}dt^{2}+N^{-2}dt^{2}+r^{2}(d\theta ^{2}+\sin ^{2}\theta d\phi
^{2})\text{,}  \label{met}
\end{equation}%
where the horizon is located at $r=r_{+}$, so $N(r_{+})=0$. We assume that a
particle having mass $m$ moves within this background with the force $F=ma$
exerted on it that causes the acceleration $a^{\mu }$, $a_{\mu }a^{\mu
}\equiv a^{2}$, by definition $a\geq 0$. We also assume that $a=a(r)$
depends on $r$ only.

Then, for four-velocity $u^{\mu }$ we have

\begin{equation}
mu^{t}=\frac{X}{N^{2}}\text{,}  \label{ut}
\end{equation}%
\begin{equation}
mu^{r}=\sigma P\text{, }P=\sqrt{X^{2}-m^{2}N^{2}}\geq 0\text{,}  \label{ur}
\end{equation}%
the factor $\sigma =\pm 1$ depending on the direction of motion,%
\begin{equation}
X=E-b(r)\text{.}  \label{X}
\end{equation}%
Here, $E$ is a constant of integration, 
\begin{equation}
b=\alpha \int_{r}^{\infty }dr^{\prime }F(r^{\prime })\text{,}  \label{beta}
\end{equation}%
$\alpha =\pm 1$. The above formulas generalize those for the RN metric.
Their derivation is quite direct and can be found in \cite{noway}, \cite{sym}%
, \cite{pk}, \cite{F}.

In the important particular case of the RN metric, taking $\alpha =+1$, $q>0$
and $Q>0$ (here, $q$ is the particle's charge, $Q$ is that of a black hole),
we have 
\begin{equation}
F(r)=\frac{qQ}{r^{2}}\text{, }b=\frac{qQ}{r}=q\varphi (r)\text{,}
\end{equation}%
where $\varphi (r)=\frac{Q}{r}$ is the Coloumb potential of a black hole.
The metric coefficient%
\begin{equation}
N^{2}=1-\frac{2M}{r}+\frac{Q^{2}}{r^{2}}\text{,}
\end{equation}%
where $M$ is a black hole mass. 

We assume that the force and acceleration caused by it decrease rapidly
enough at infinity. Then, $E$ has the usual meaning of the energy at
infinity. The forward-in-time condition $u^{t}>0$ requires 
\begin{equation}
X\geq 0.  \label{forw}
\end{equation}%
Let particles 1 and 2 move from infinity and collide in some point $r_{c}$.
The energy $m_{0}$ in the center of mass frame%
\begin{equation}
m_{0}^{2}=-(m_{1}u_{1}^{\mu }+m_{2}u_{2}^{\mu })(m_{1}u_{1\mu }+m_{2}u_{2\mu
})=m_{1}^{2}+m_{2}^{2}+2m_{1}m_{2}\gamma \text{,}  \label{cm}
\end{equation}%
where $\gamma =-u_{1\mu }u^{2\mu }$ is the Lorentz factor of relative
motion. It follows from the above equations that%
\begin{equation}
\gamma =\frac{X_{1}X_{2}-\sigma _{1}\sigma _{2}P_{1}P_{2}}{m_{1}m_{2}N^{2}}.
\label{lor}
\end{equation}

This is just the point where particle dynamics described by eqs. (\ref{ut})
- (\ref{beta}) reveals itself.

One reservation is in order. We assume that the background is described by
the spherically symmetric metric (\ref{met}). In doing so, backreaciton of a
particle and external sources on the metric is neglected. For the case of
the Schwarzschild metric this implies that $m\ll M$, where $M$ is a black
hole mass.

\section{Scenarios of collision}

We consider the case $\alpha =+1$, $\sigma _{1}=\sigma _{2}=-1$. In
particular, for the RN black hole $a=\frac{\left\vert qQ\right\vert }{mr^{2}}
$. However, we may leave a general $a(r)$ not specifying it. In particular,
one can consider motion of particles in the Schwarzschild background under
the action of some force \cite{pk}, \cite{pk2}. This can also lead to the
BSW effect \cite{F}. Now we will see that this also admits the SPP.

Let us consider reaction $1+2\rightarrow 3+4$. For simplicity, we choose the
case of pure radial motion of all particles. Then, it follows from the
conservation laws of energy and radial momentum that (see Sec. IV of \cite%
{center})

\begin{equation}
X_{3}=\frac{1}{2m_{0}^{2}}(X_{0}\Delta _{+}-P_{0}\sqrt{\Delta
_{+}^{2}-4m_{0}^{2}m_{3}^{2}})_{c},  \label{x3}
\end{equation}%
\begin{equation}
X_{4}=\frac{1}{2m_{0}^{2}}(X_{0}\Delta _{-}+P_{0}\sqrt{\Delta
_{+}^{2}-4m_{0}^{2}m_{3}^{2}})_{c},
\end{equation}%
\begin{equation}
\Delta _{\pm }=m_{0}^{2}\pm (m_{3}^{2}-m_{4}^{2}),
\end{equation}%
\begin{equation}
X_{i}=E_{i}-b_{i}(r),  \label{xb}
\end{equation}%
\begin{equation}
P_{i}=\sqrt{X_{i}^{2}-m_{i}^{2}N^{2}}.  \label{p}
\end{equation}%
Here, the integer $i$ runs from $0$ to $4$, $\sigma _{i}=\pm 1$, subscript
"c" refers to the point of collision$,X_{0}=X_{1}+X_{2}$, $m_{0}$

The outcome of collision depends strongly on the relation between particles'
parameters, say the energy and charge in the RN case. Then, following a
standard terminology, we can classify all particles depending on $X(r_{+})$.
If $X(r_{+})>0$ is separated from zero, a particle is called usual.\ If $%
X(r_{+})=0,$ a particle is called critical. If $X(r_{+})=O(N_{c})$, where $%
N_{c}\ll 1$, a particle is called near-critical.

We assume that particle 1 is near-critical and particle 2 is usual. More
precisely, we specify deviation from the criticality in the form%
\begin{equation}
b_{1}(r_{+})=E_{1}(1+\delta _{1})  \label{b1}
\end{equation}%
where%
\begin{equation}
\delta _{1}=C_{1}N_{c}+O(N_{c}^{2})  \label{d1}
\end{equation}%
with $C_{1}<0$ and $N_{c}\ll 1$. We also assume the validity of the Taylor
expansion in the form%
\begin{equation}
b_{1}(r_{c})\approx b_{1}(r_{+})+A(r_{c}-r_{+})\text{,}  \label{bA}
\end{equation}%
where $A$ is come constant. The nonextremal nature of a black hole comes
into play just here: for nonextremal black holes $N_{c}^{2}\sim r_{c}-r_{+}$%
. Meanwhile, the corrections to $X_{1}$ (\ref{xb}) have two contributions.
The first one stems from the second term in (\ref{bA}). The second
contribution arises due to the correction (\ref{b1}) to the critical value $%
b_{1}(r_{+})=E_{1}$ because of $\delta _{1}$. We see that the first type of
corrections is negligible having the order $N^{2}$. Thus in the vicinity of
the horizon%
\begin{equation}
X_{1}=E_{1}\left\vert C_{1}\right\vert N_{c}+O(N^{2})\text{.}  \label{x1}
\end{equation}

Also, for near-horizon collisions, $X_{0}\approx X_{2}$, $P_{0}\approx X_{0}-%
\frac{m_{0}^{2}N^{2}}{2X_{0}}$ and all nonzero quantities can be taken on
the horizon instead of $r_{c}.$ It is seen from (\ref{cm}), (\ref{lor}) that 
\begin{equation}
m_{0}^{2}\approx \frac{2\left( X_{2}\right) _{+}z}{N_{c}}\approx \text{ }%
\Delta _{+}\text{,}  \label{md}
\end{equation}%
where%
\begin{equation}
z=E_{1}C_{1}-\sqrt{E_{1}^{2}C_{1}^{2}-m_{1}^{2}}\text{.}  \label{z}
\end{equation}

Collecting all terms carefully, one obtains from (\ref{x3}) after simple
algebra that

\begin{equation}
X_{3}\approx \frac{N_{c}}{2}(z+\frac{m_{3}^{2}}{z})\text{.}  \label{Xz}
\end{equation}

We would like to remind a reader that eq. (\ref{x3}) from which (\ref{Xz})
is obtained, is direct consequence of particle dynamics. More precisely, it
follows from the conservation of energy and radial momentum. This was shown
in Sec. IV of Ref. \cite{center} where a reader can find the details. In
addition to particle dynamics, we used the proximity to the critical state
for particle 1 because of which the approximate expression (\ref{x1}) was
obtained from (\ref{xb}), (\ref{b1}), (\ref{bA}) and inserted in eq. (\ref%
{x3}).

Thus particle 3 turns out to be near-critical. From the other hand, for a
particle of such a type the approximate expression similar to (\ref{x1})
should be valid, so 
\begin{equation}
X_{3}\approx \left\vert C_{3}\right\vert E_{3}N_{c}\text{,}  \label{XC}
\end{equation}%
where $\delta _{3}=C_{3}N\,$\ and $C_{3}<0$ controls relationship between
parameters of particle 3.

Say, for the RN metric, $b=q\varphi $, where $\varphi =\frac{Q}{r}$ is the
electric potential of a black hole, $Q$ being its charge. Then, for the
exactly critical particle 
\begin{equation}
E=q\varphi (r_{+}),  \label{crit}
\end{equation}%
so $q=\frac{Er_{+}}{Q}$, for the nonextremal black hole $r_{+}>Q$. For
near-critical particle 3, we can choose

\begin{equation}
q_{3}=E_{3}\frac{r_{+}}{Q}(1+\delta _{3})\text{.}  \label{q3}
\end{equation}%
By substitution in (\ref{X}) and discarding the terms $O(N^{2})$, we see
that (\ref{XC}) does hold true.

We would like to stress that, had we taken two neutral particles, we would
have obtained collision of two usual particles. Then, according to general
rules \cite{ban}, \cite{jl}, \cite{prd}, the BSW effect would be impossible.
The Penrose process would be impossible in this situation as well since for
neutral particles there are no negative energies. Thus the electric charge
is a necessary ingredient of a process. For\ a more general situation, it is
necessary that one particle move on the action of a force while another one
can be free.

Comparing (\ref{Xz}) and (\ref{XC}), we find that 
\begin{equation}
E_{3}\approx \frac{1}{2\left\vert C_{3}\right\vert }(z+\frac{m_{3}^{2}}{z})%
\text{.}  \label{EC}
\end{equation}

If we choose $C_{3}\rightarrow 0$, then formally $E_{3}\rightarrow \infty $
becomes unbounded, as it should be for the SPP. Thus we see that for
nonextremal black holes, the proximity to the criticality requires not only
the validity of expansion (\ref{d1}) for particle 3, like it was in the
extremal case \cite{rn}. Additionally, the coefficient $C_{3}$ has to be
small.

This is not the end of story. We must also \ check that the scenario under
consideration is able to describe the process when particle 3 escapes to
infinity instead of fall in a black hole. This requires that $\sigma _{3}=+1$%
. (The particle with $\sigma _{3}=-1$ simply falls in a black hole in
contrast to the extremal case where it can bounce from the potential barrier
and escape \cite{rn}.) The list of possible scenarios is given in eqs. 29,
30 of \cite{center}. It is shown there that the aforementioned condition
requires 
\begin{equation}
\Delta _{+}N_{c}-2m_{3}X_{0}>0\text{.}  \label{esc}
\end{equation}%
In the main approximation using (\ref{md}), (\ref{z}) we have from (\ref{esc}%
) that%
\begin{equation}
m_{3}<z<m_{1}\text{.}
\end{equation}

Thus, although there is no upper bound on $E_{3}$, there is such a bound on $%
m_{3}$. In doing so, one obtains from (\ref{p}) that%
\begin{equation}
P_{3}\approx \frac{N}{2}(z-\frac{m_{3}^{2}}{z}).
\end{equation}

\section{Choice of initial state}

The formulas written above are valid for the process as such, independently
of the origination of the initial state. This implies the collision of one
near-critical and one usual particles that were created somehow in the
vicinity of the horizon. Meanwhile, the most physically interesting
situation arises when particles 1 and 2 come from infinity. However, in this
respect severe restrictions are relevant since in general the critical or
near-critical particle cannot overcome the potential barrier in the
nonextremal black hole background \cite{gp}, \cite{prd}, \cite{gao}, \cite%
{kras}. Happily, there are two particular cases when this becomes possible.
(i) The nonextremal black hole is close to extremality, its surface gravity $%
\kappa $ is small. (ii) Both particles 1 and 2 have big energy from the very
beginning. This represents the so-called "energy feeding problem", its
analogue for extremal black holes was discussed in Sec. IV C1 of Ref. \cite%
{axis}. However, it was shown in Sec. VII of \cite{can} that, although the
initial energy should be big, the outcome due to collisions of
ultrarelativistic particle can give significant relative gain in energy $%
m_{0}$.

In the present work, as a matter of fact, we showed that significant gain in
energy $E_{3}$ is possible as well. By itself, it does not depend on an
initial energy at all, its value being controlled by the deviation from
criticality measured by the quantity $\left\vert C_{3}\right\vert $. For the
RN metric, the electric charge of debris at infinity is defined according to
to (\ref{q3}), (\ref{EC}). Thus, big initial energies of incoming particles
do not depreciate such a scenario since the outcome is more "profitable"
than income anyway. Cases (i) and (ii) are discussed in detail in \cite{can}%
, so we do not repeat their details here. What is important for us is that
at least with some additional requirements, scenarios of high energy
collisions of particles 1 and 2 for nonextremal black holes work suggest a
suitable initial state, so the present scenario describing behavior of
particles 3 and 4 also works.

\section{Extremal black hole versus nonextremal one}

It is instructive to compare the results with those for the extremal RN
black hole. Then, 
\begin{equation}
X_{3}=-E\delta +EN+O(N^{2}),  \label{xet}
\end{equation}%
\begin{equation}
X_{3}(N_{c})\approx (1-C_{3})E_{3}N_{c}  \label{xetn}
\end{equation}%
- see eq. (19) in \cite{rn}. Therefore, the limit $C_{3}\rightarrow 0$ is of
no use in that case and more refined analysis was required there, with
account of all possible $C_{3}$ (positive, negative and zero). It was shown
that there are different scenarios, and only the ones with $C_{3}\geq 0$
give rise to the SPP process \cite{rn}. In doing so, there are two options.
If immediately after collision particle 3 moves to infinity, there is no
upper bound on $E_{3}$ but there exists such a bound on $m_{3}$. If it moves
towards a black hole, bounces from the barrier and escapes to infinity,
there is an upper bound neither for $E_{3}$, nor for $m_{3}$. For the
nonextremal case, the relevant scenario, as we saw, is more simple, eq. (\ref%
{x1}), (\ref{XC}) are valid instead of (\ref{xet}), (\ref{xetn}). There is
no turning point, particle 3 relevant for te SPP, moves to infinity at once
after collision.

The difference between the extremal and nonextremal cases reveals itself
also in potential realization of the SPP\ in a real world. For the extremal
case, let us assume that critical particle 1 falls from inifnity with $%
E_{1}=q_{1}$ according to (\ref{crit}) since $\varphi (r_{+})=1$. If
initially the particle was at rest at inifnity, $E_{1}=m_{1}$. Taking $%
q_{1}=\left\vert e\right\vert $ where $e$ is the elementary charge, one
obtains that $m_{1}\approx 10^{-6}$ g$,$ so one deals with a macroscopic
object instead of an elementary particle that was pointed out in Sec. VII of 
\cite{nem}. Meanwhile, for nonextremal black holes, there are two changes.
First, an initial particle should be ultrarelativistic (unless the black
hole is very closed to the extremal state) - see \cite{can}. Thus $%
E_{1}=m_{1}\gamma $, where $\gamma \gg 1$ is the corresponding Lorentz
factor. Second, now $\varphi (r_{+})\sim \frac{Q}{M}$. Thus (\ref{crit})
reads $m_{1}\sim \frac{Q}{\gamma M}q_{1}$. If $\frac{Q}{M}\lesssim 1$, this
means that $m_{1}\sim \frac{q_{1}}{\gamma }$. Taking, again, $%
q_{1}=\left\vert e\right\vert $, we obtain that $m_{1}\approx \frac{10^{-6}}{%
\gamma }$g$.$ If, say, $\gamma =100\,$, $m_{1}\approx 10^{-4}$g$,$ so an
object can be still considered as macroscopic instead of being an elementary
particle. Formally, for $m_{1}$ to become equal to the electron mass, one
must have $\gamma \approx 10^{21}$ but such tremendous initial velocities
are unphysical. Thus the main conclusion made in Sec. VII of \cite{nem} that
instead of elementary particles, macroscopic objects are involved in the
process under discussion, retains its validity.

Until recently, there was general belief that the electric charge of black
hole is absent or completely negligible. Meanwhile, there are some indirect
indications that, although being extremely small, such a charge can differ
from zero \cite{galac}. It is reasonable to require that $m>m_{e}$ where $%
m_{e}$ is the electron mass. Then, we obtain that $\frac{Q}{M}>\frac{\gamma
m_{e}}{\left\vert e\right\vert }$, whence $\frac{Q}{M}>10^{-22}\gamma $.
Thus for, say, $\gamma =100$ we obtain a lower bound on the charge $\frac{Q}{%
M}>10^{-22}$ for the realizaiton of the SPP for electrons. For protons as
critical particles, one must require $\frac{Q}{M}>10^{-19}$.

The above discussion does not affect the condition $E_{3}\gg E_{1}$ and the
corresponding energy gain since both critical energies are proportional to $%
\varphi (r_{+})$, so $\varphi (r_{+})$ drops out and $\frac{E_{3}}{E_{1}}=%
\frac{q_{3}}{q_{1}}$. If, say, we take $q_{1}=\left\vert e\right\vert $ and $%
q_{3}=$ $Z\left\vert e\right\vert $, the bound $Z\leq 170$ follows from
relaistic estimates that envolve stable nuclei \cite{zp}. Meanwhile, this is
compatible with the relations $Z\gg 1$, $E_{3}\gg E_{1}$, so the surplus is
quite significant. Also, we would like to point out that, in principle, the
obtained results apply to the case when the charge is not electric one but
is tidal. This can be formed as an effective charge, being imprint of
high-dimensional space-times in our four-dimensional one, having potential
astrophysical consequences \cite{nar}, see also \cite{zak} and references
therein.

\section{Discussion and conclusions}

We showed that a nonextremal static black hole is pertinent to the SPP. In
doing so, there is a restriction on a mass escaping to infinity but there is
no upper bound on its energy. (This holds true as long as backreaction is
negligible, so test particle approximation is valid.) Thus the SPP exists
both for the extremal and nonextremal RN black hole. Moreover, all this
consideration applies even to the Schwarzschild black hole, provided some
force is exerted on the critical particle. This includes both any external
force or electric repulsive (since $b\sim qQ>0$ is required) force in the RN
metric, if this force is small enough. The latter means that the force does
not changes background significantly (that remains approximately
Schwarzschildean) but affects particle's motion. It is worth noting that the
latter situation occurs also in a quite different context when the
Schwarzschild black hole is immersed in a weak magnetic field that can lead
to high energy collisions \cite{fr}.

Usually, the electric charge is similar to rotation in black hole physics in
many aspects. In particular, this concerns the BSW effect \cite{ban}, \cite%
{jl}. However, now this similarity breaks down since the SPP does not exist
for rotating black holes but is possible for static charged ones. Now, this
is seen not only for extremal black holes \cite{rn}, \cite{axis} but also
for nonextremal ones.

In the present context, we would like to make a technical but important
remark. As we saw above (see discussion around eq. (\ref{x1})), the
difference between nonextremal and extremal black holes manifests itself in
the different role of corrections that come from the force for the escaping
particle. In the extremal case, they have the same order as the term due to
the deviation of parameters from the critical state. In the nonextremal one,
they are negligible and this simplifies consideration greatly.

The key difference between rotating and static black holes and in the
context under discussion lies in the role of the centrifugal barrier. One of
two new particles after collision should be near-critical. But this barrier
the critical particle with very high energy to reach infinity after
collision in the first case that destroys the SPP both in the extremal and
nonextremal cases (see Sec. VII in \cite{center}). However, for static black
holes (say, the RN one) there is no such a barrier at all. The results of
the present work extends essentially the area of validity of high energy
processes since astrophysically relevant black holes are nonextremal. It is
of interest to understand, how the effects of such a kind can reveal
themselves in the accretion discs around black holes and for the charge
which is not electric but tidal. This needs separate treatment.

This work is performed according to the Russian Government Program of
Competitive Growth of Kazan Federal University.

\end{document}